%% file: main.tex
\documentclass{class}
\usepackage{graphicx}
\usepackage{epstopdf, epsfig}
\usepackage{siunitx}
\usepackage{amssymb} 
\usepackage{dirtytalk}
\usepackage{textcomp}

\usepackage{newtxtext}
\usepackage{newtxmath}
\title{Reactive experimental control of turbulent jets}
\author{Diego B. S. Audiffred\aff{1},
  André V. G. Cavalieri\aff{1},
  Igor A. Maia\aff{1},
  Eduardo Martini\aff{2}
  \and Peter Jordan\aff{2}}

\affiliation{\aff{1} Instituto Tecnológico de Aeronáutica, São José dos Campos, São Paulo, Brazil
\aff{2}Institut Pprime-CNRS-Université de Poitiers-ENSMA, Chasseneuil du Poitou, Nouvelle-Aquitaine, France}

\begin{document}
\maketitle
\section*{Abstract}
\begin{abstract}
We present an experimental study of reactive control of turbulent jets. We target axisymmetric disturbances associated with coherent structures, which are known to underpin the peak sound radiation of turbulent jets. We first consider a forced jet flow case, such that the coherent structures can be amplified above background levels, which makes it easier to detect them by the sensors. We then consider the more challenging case of a natural jet, i.e., without artificial forcing. The control strategy explores linear convective mechanisms in the initial jet region, which justifies application of linear control theory. The control law is constructed in the frequency domain, based on empirically determined transfer functions. The Wiener-Hopf formalism is used to enforce causality, providing an optimal causal solution, thus, preventing the drop in performance that may be observed in flow control applications that use simpler wave-cancellation methods. With this approach, we could improve the control performance of forced turbulent jets compared to results obtained in previous studies, attaining order-of-magnitude attenuation of power spectra of velocity fluctuations. Furthermore, we could obtain substantial levels of attenuation of natural turbulent jets, of about 60\% in power spectra for the most amplified frequencies. These results open new directions for the control of turbulent flows.
\end{abstract}

\input{sec_Intro}
\input{sec_Exp}
\input{sec_CtrlDesign}
\input{sec_Results}

\input{sec_Conclusion}

\section{Acknowledgements}
This work has received funding from the European Union’s Horizon 2020 research and innovation programme under grant agreement No861438. We also wish to acknowledge the financial support from the Coordination for the Improvement of Higher Education Personnel (CAPES) and from the São Paulo Research Foundation (FAPESP), grants 2022/03279-5 and 2019/26546-6. We also acknowledge the technical support provided by Anton Lebedev, who greatly assisted the experiments conducted for this work.

\bibliographystyle{style}
\bibliography{literature}

\end{document}

%% file: sec_Intro.tex
\section{Introduction}
\label{sec:intro}

Attenuation of jet exhaust noise is one of the main concerns of the aircraft industry. Since the introduction of jet engines, many studies have been conducted aiming at reducing jet exhaust noise \citep{Wirt1966, Seiner1989, Saiyed2000, Ginevsky2004, Morris2019, zigunov2022}. However, jet noise still remains one of the main noise sources in an aircraft, and is dominant during takeoff \citep{Leylekian2014,Huff2016}. Furthermore, the Advisory Council for Aeronautics Research in Europe (ACARE) has established a goal of reducing the perceived noise emission of flying aircraft by 65\% in 2050 with respect to the values observed in the year 2000 \citep{acare}. 

Jet noise generation is mainly caused by the high-velocity exhaust gases that leave the nozzle and meet the atmospheric air, which creates turbulence with a wide range of turbulent structures \citep{suzuki2006}. Among these, there are large coherent flow structures \citep{mollo1967, crow_champagne_1971, LAU1972, moore1977}, which are acoustically efficient and dominate the downstream sound radiation \citep{Tam1996, tam_2008, MICHALKE19831}. These coherent structures, also known as wavepackets, are underpinned by the Kelvin-Helmholtz instability, and are characterized by a large spatial envelope of growth and decay, and a high degree of organization \citep{gudmundsson_colonius_2011, PeterTim, cavalieri2019wave}.


When it comes to reducing jet noise, lowering the jet velocity may be seen as the obvious thing to do, since the total acoustic power is roughly proportional to $U^8$, as predicted by \citet{Lighthill1952}. In the last decades, larger engines were developed to increase efficiency. A consequence of a larger turbine is that, for a given thrust, the jet velocity can be reduced,  also reducing the radiated noise. However this strategy is reaching its limits due to structural and ground clearance reasons. Therefore, to further reduce jet noise, it is important to explore new solution strategies \citep{Leylekian2014}. Another well-known noise-mitigation strategy consists in using modified nozzle geometries, such as chevron nozzles, which has been shown as an effective technology to reduce jet noise \citep{Tide2009, Huff2007}. The mechanism behind the low-frequency noise reduction observed with chevron nozzles is the increase in streamwise vorticity in the jet shear layer. This enhances the mixing, reduces the potential core length as has a stabilizing effect on Kelvin-Helmholtz wavepackets\citep{Alkislar2007, sinha2016, lajus2019}. One advantage of such passive control devices is that they do not require an energy input. However, it typically leads to lower thrust efficiency. Another strategy is to use reactive flow control, where real-time sensor readings are used in order to build a control law and generate an unsteady actuation, which responds in real-time to unsteady fluctuations in the flow.

Since the main component responsible for jet noise generation are wavepackets, whose underlying dynamics can be modeled using linear models, especially near the nozzle exit \citep{cavalieri2013}, the application of linear control theory is a suitable candidate for designing an explicit control law for jet noise attenuation. The first explicit attempt of actively controlling instability waves in jets was investigated in a 2D model of a nozzle edge, where the suppression of a time-harmonic instability wave was shown to be theoretically possible via an external acoustic excitation with properly chosen amplitude and phase \citep{Kopev2008}. Few years later, this was performed experimentally, where harmonically forced disturbances in a turbulent jet have been attenuated by an external acoustic source \citep{Kopiev2013} and also with plasma actuators \citep{Kopiev2014}. This study evolved to the development of a strategy for controlling natural instability waves in subsonic turbulent jets \citep{Belyaev2018}, which was later tested experimentally in a feedback control scheme \citep{faranosov2019}, where axisymmetric instability waves of a natural jet were attenuated using a plasma actuator placed inside the nozzle, near the exit. For such case, the control was only aimed at a Strouhal number of $St = 0.46$, and attenuation was achieved for a narrow frequency range around this target frequency. More recently, the Inverse Feed-forward Control (IFFC) method \citep{Sasaki2018} was used in experimental control studies of band-limited stochastically forced turbulent jets, with the objective of canceling axisymmetric hydrodynamic wavepackets using synthetic jets actuators\citep{Maia2021,Maia2022}. Several studies were also developed focusing on the control of mixing layers, which can be seen as a simple model for the initial region of the jet shear layer \citep{WEI2005, Parezanovi2014, Shaqarin2018}. 

The IFFC method opened the possibility of controlling disturbances in a broader spectrum, which was an important step towards the control of a turbulent jet. The IFFC approach is constructed in the frequency domain. When converted back to the time domain, this approach casts the control law as a convolution between sensor readings and a control kernel. In general, this convolution is non causal, i.e., it requires future information to construct the present flow actuation. Real-time applications require that only the causal part of the kernel (present and past measurements) be used. In IFFC, this is enforced by truncating the kernel to its causal part, which might result in a substantial drop in performance \citep{Brito2021}. This issue can be circumvented with the Wiener-Hopf formalism, which provides a framework to compute an optimal causal control kernel \citep{Martinelli2009, Martini2022}. Recent numerical and experimental studies have shown that Wiener-Hopf-based control can improve on the results of IFFC for amplifier flows \citep{Martini2022, Audiffred2023}. This suggests that the turbulent jet control can benefit from the Wiener-Hopf technique as well.

In the present work we aim at attenuating axisymmetric hydrodynamic wavepackets with the application of the Wiener-Hopf technique in a turbulent jet. We first consider the case of a forced jet. The application of an external forcing increases the amplitudes of the targeted flow structures, making it easier to identify and control them \citep{crow_champagne_1971, moore1977, gudmundsson_colonius_2011}. This is the same underlying idea of the work by \citet{Maia2021}. This avoids some of the difficulties of the control of a natural jet, and is a necessary first step to test the control strategy and provide a proof of concept of its suitability. Furthermore, we emphasise, that, as  in \citet{Maia2021, Maia2022}, forcing amplitudes are kept small enough so that the jet responds \textit{linearly}. This makes sure that the instability mechanisms in the forced jet are essentially the same as in their natural counterparts. In the second part of the study, we extend the control approach to the natural jet case. As highlighted above, this is a more challenging scenario, due to the difficulty in sensing the low-energy axisymmetric wavepackets. A natural jet case is a condition of greater interest since turbulent jets in practice do not involve an explicit forcing.

The characteristics of the control experiment is discussed in the next section. In \S \ref{WHcontrole}, the Wiener-Hopf technique is briefly presented, along with the control design considering both the Wiener-Hopf and the inverse feed-forward control methods. Then, the results obtained by each of these methods are compared and discussed in \S \ref{PrelResults}. Further discussion about the control applied and the results obtained are offered in \S \ref{conclusion}, which concludes the paper.





%% file: sec_Exp.tex
\section{Experimental Setup}\label{experimentos}

\begin{figure}
	\centering
    \includegraphics[width=0.7\textwidth]{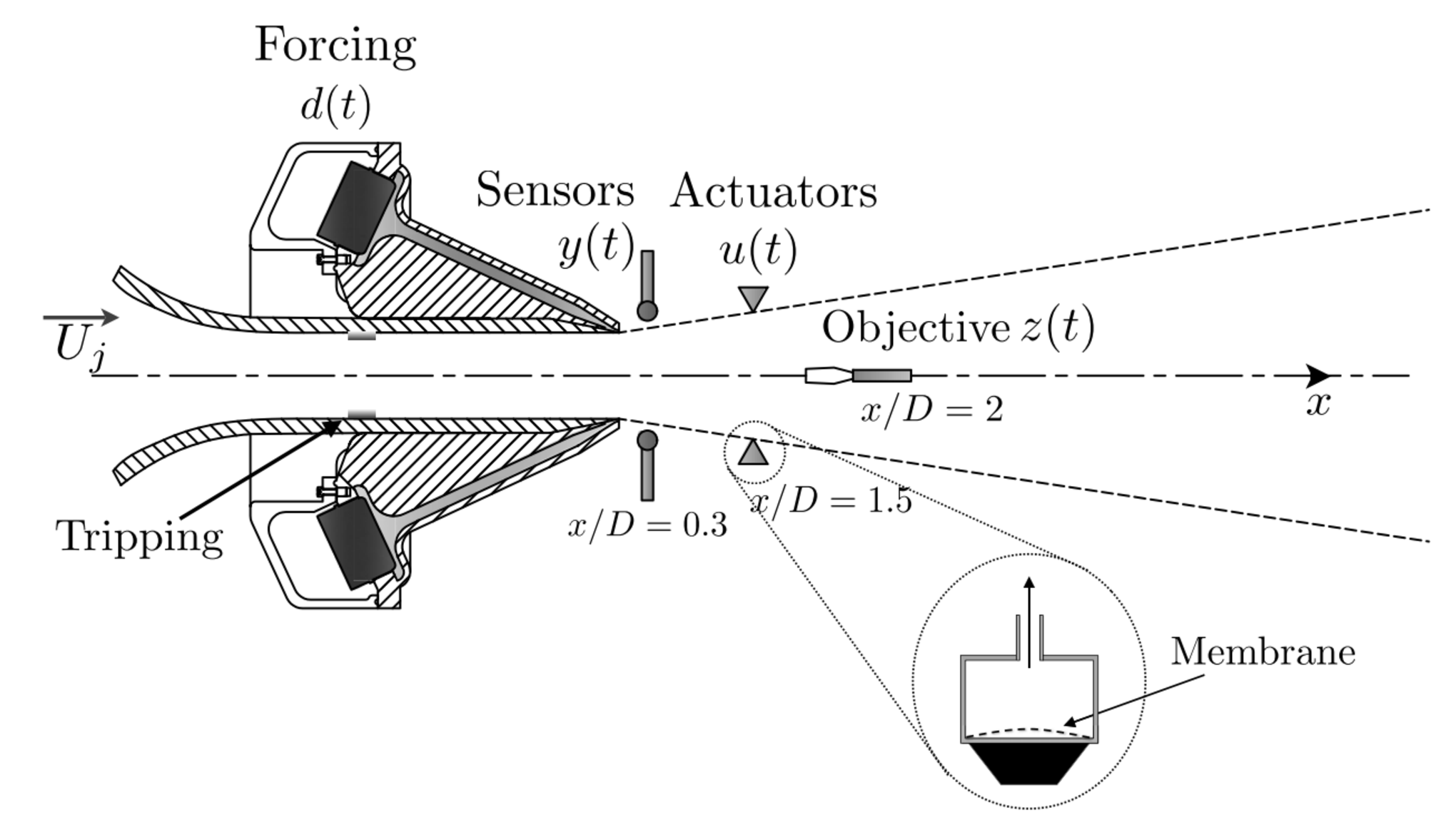}
    \includegraphics[width=0.7\textwidth]{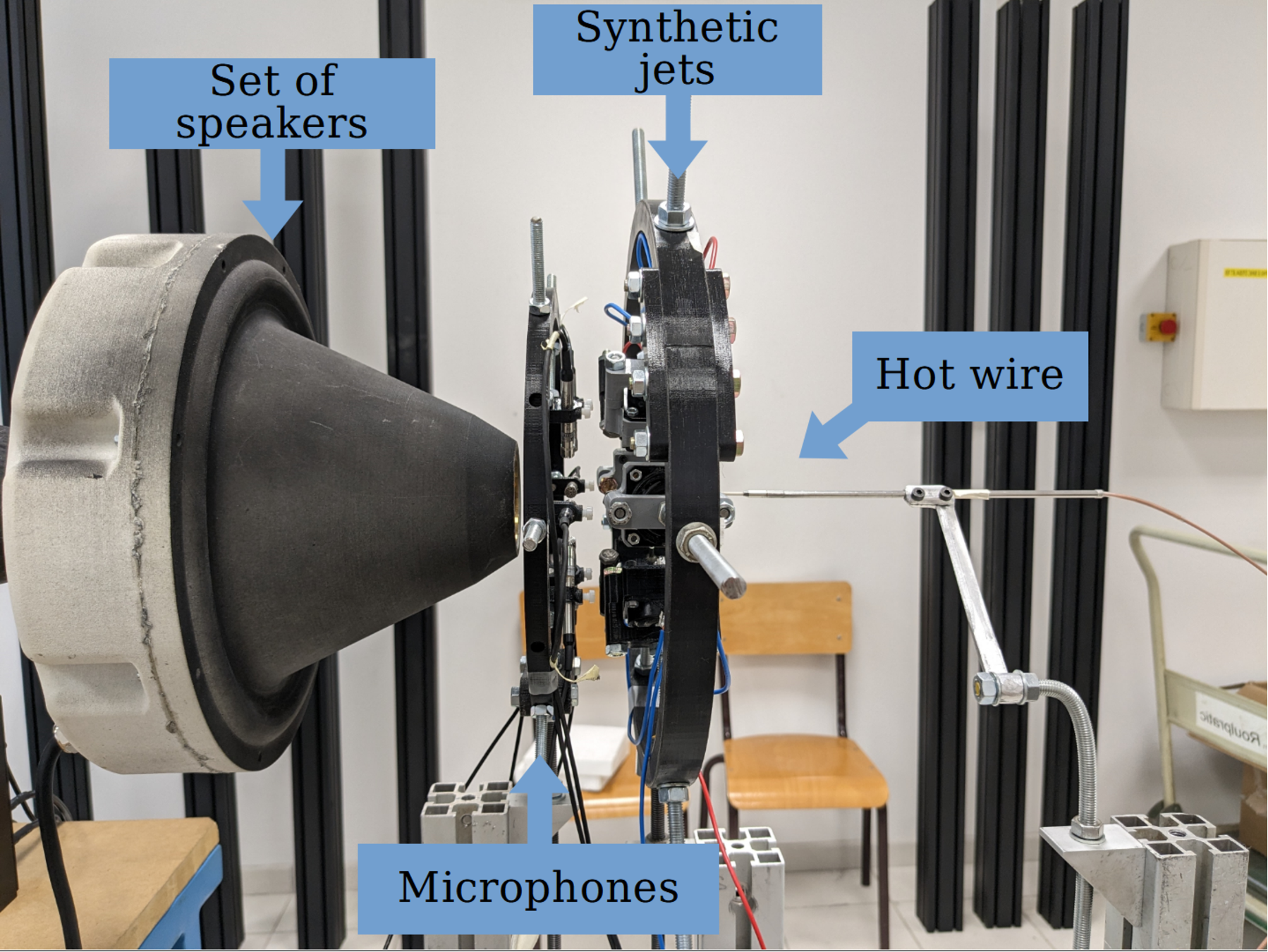} \\
	\caption{Sketch of the control scheme of the experimental setup, showing sensors and actuators positioning \citep{Maia2021}(top), and a side view photo of the control setup (bottom).}
	\label{fig:expsetup}
\end{figure}

\begin{figure}
	\centering
	\includegraphics[width=0.7\textwidth]{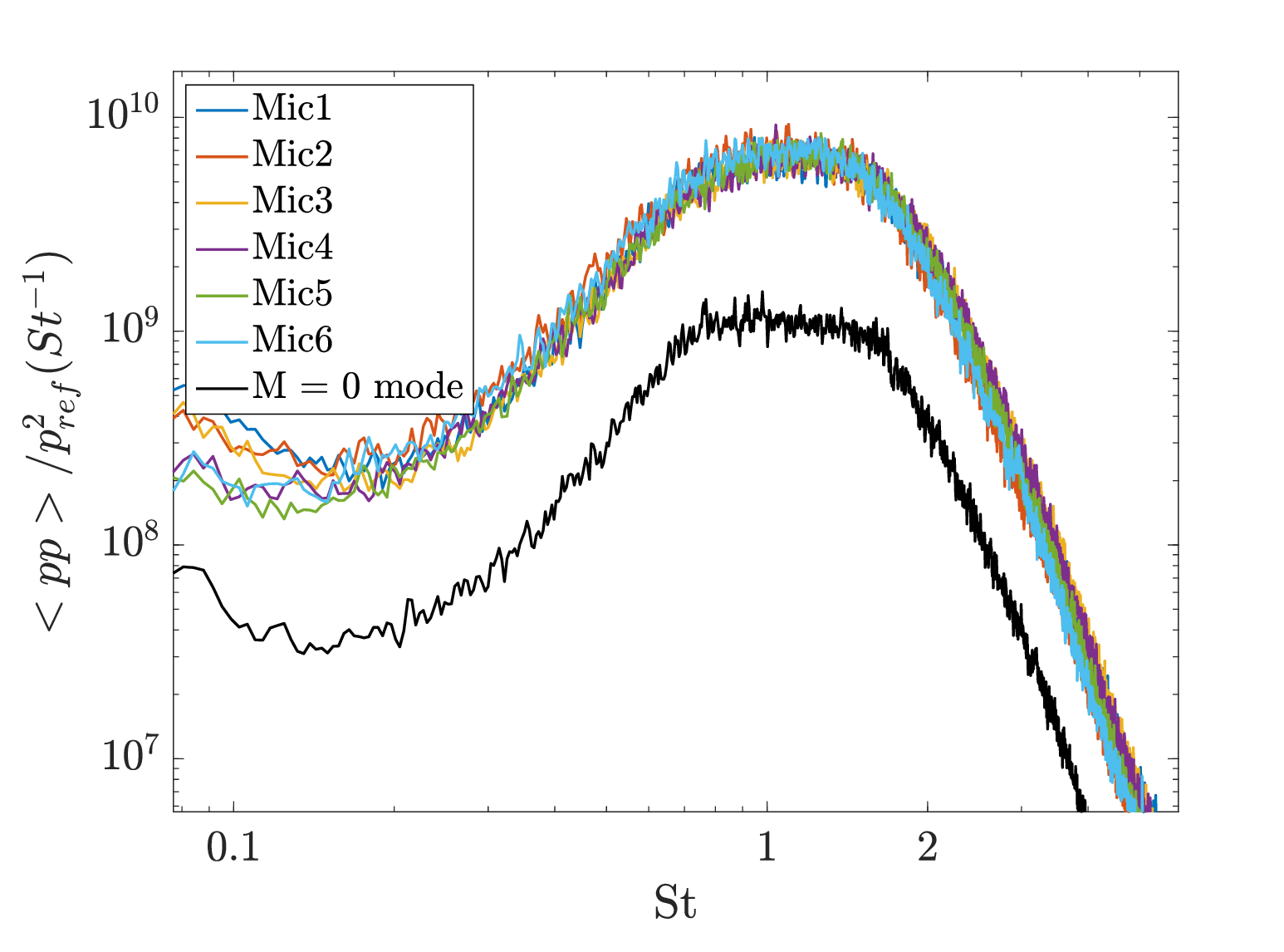} \\
	\caption{Pressure spectra measured by the six microphones and the resulting m = 0 mode. Measurements taken at $x/D = 0.3$ and $r/D = 0.55$.}
	\label{fig:m0mode}
\end{figure}

Real-time flow control experiments of turbulent jets have been performed at the Pprime Institute. We consider jets at Mach number of $0.05$ and a Reynolds number of $Re  = 5 \times 10^4$. Transition to a fully turbulent jet was guaranteed by a strip of carborundum particles placed $2.5D$ upstream of the nozzle exit plane, where $D$ is the nozzle diameter, which is $50$ mm. The experimental setup was based on the experiments reported by \citet{Maia2021}, which is represented in fig. \ref{fig:expsetup}. It consisted of six microphones, working as pressure sensors, providing the axisymmetric pressure mode information, which is the input signal, $y$, for the control law. The axisysmmetric pressure mode obtained for the unforced jet is shown in fig. \ref{fig:m0mode}. Spectra of the six microphones are very close to each other, indicating azimuthal homogeneity. The axisymmetric mode has only a small fraction of the power, but it is known to be related to the peak acoustic radiation of subsonic jets \citep{juve1979, cavalieri2013}.

The actuation, $u$, was provided by synthetic jets generated by a set of six speakers. These actuators primarily work through a blowing and suction process of zero-net mass flux that generate a train of vortices that moves away from the device's orifice due to self induced velocity \citep{wang_feng_2018}. Further downstream,  on the jet center line,  a hot wire has been used to take streamwise velocity measurements. The microphones were placed at a distance $x/D = 0.3$ from the nozzle exit, the actuators at a distance $x/D = 1.5$, and the  hot wire at $x/D = 2$. The radial distance of the microphones from the jet center line was $r/D = 0.55$, while the actuators were placed at a radial distance $r/D = 0.7$. These choices of positioning avoided intrusion of sensors and actuators in the jet plume, remaining nonetheless in the near field. The first set of control experiments was performed considering the application of an axisymmetric forcing, $d$, provided by synthetic jets generated by a set of eight speakers. The application of a forcing signal helps improving the coherence between sensors and actuators, and also allows us to compare some of the results obtained here with those obtained by \citet{Maia2021}. 

 For the forcing, we have considered two configuration of band-limited stochastic signals, one in a Strouhal number range $0.3 \leqslant St \leqslant 0.45$, and the other for Strouhal numbers in the interval $0.3 \leqslant St \leqslant 0.85$. The actuation was designed in the same Strouhal number range as the forcing, as will be described in the next section. In the case of the unforced (natural) jet, the actuation is within the band $0.3 \leqslant St \leqslant 0.85$. This range corresponds, to a great extent, to the most amplified frequencies at the sensor position, due to the Kelvin-Helmholtz mechanism \citep{Maia2021, Maia2022}. Further downstream, at the control target, growth rates are smaller, but the jet still remains convectively unstable up to $St = 0.65$ \citep{Maia2021}. Therefore, the frequency range considered for the forcing falls within the unstable region of the spectrum. This is one of the ranges considered by \citet{Maia2021}.

The Simulink tool, which is integrated with MATLAB, has been used to build the control model that was employed in the ControlDesk software from dSPACE, and the MicroLabBox system acquisition from dSPACE has been used for the real-time sensor readings and actuation, following the same procedure as earlier flow control works of our group \citep{Brito2021, Audiffred2023}. 

%% file: sec_CtrlDesign.tex
\section{Control law design}\label{WHcontrole}

\subsection{Overview of the Wiener-Hopf technique}\label{WienerHopf}

The Wiener-Hopf technique \citep{Noble_1958, Martini2022} is applied for the jet control problem, and its performance is compared to that of a wave-cancelling approach \citep{Sasaki2018, Brito2021, Maia2021}. The same control parameters and transfer functions are used for both methods. In this section, the Wiener-Hopf technique is briefly presented, and further details about this method and its application to control problems can be found in \citet{Martini2022}. Wiener-Hopf equations are typically associated with problems where a half domain condition needs to be satisfied, such as a half-convolution problem \citep{Noble_1958,Martini2022},

\begin{equation}
   \int_{0}^{\infty} \mathbf{\hat{S}}(t-\tau)\mathbf{{W}}_+(\tau) d\tau= \mathbf{T}(t),  \hspace{1cm} t > 0.
    \label{eq:WHf}
\end{equation}

\noindent where $\mathbf{\hat{S}}$ and $\mathbf{T}$ are known functions and $\mathbf{{W}}$ is a unknown function (kernel). Extending this equation for $t < 0$, and taking a Fourier transform, leads to the following equation in frequency domain:

\begin{equation}
   \mathbf{\hat{S}}(\omega)\mathbf{\hat{W}}_+(\omega) = \mathbf{\hat{W}}_-(\omega)+\mathbf{\hat{T}(\omega)},
    \label{eq:WHf22}
\end{equation}

\noindent where + and - subscripts indicate that the function is analytical in the upper and lower complex half planes, respectively. This is ensured for $\mathbf{{W}}_+$ and $\mathbf{{W}}_-$ as these functions are defined to be zero for negative and positive times, respectively. $\mathbf{{W}}_+$ and $\mathbf{{W}}_-$ may also be referred to as causal and anticausal functions, respectively; these definitions carry over to all $+$ and $-$ functions.

Solution of such problem can be achieved with multiplicative and additive factorization,

\begin{equation}
   \mathbf{\hat{S}}(\omega)= \mathbf{\hat{S}_-}(\omega)\mathbf{\hat{S}_+}(\omega),
    \label{eq:Hfac}
\end{equation}

\begin{equation}
   ( \mathbf{\hat{S}}_-^{-1}(\omega)  \mathbf{\hat{T}}(\omega))  = ( \mathbf{\hat{S}}_-^{-1}(\omega)  \mathbf{\hat{T}}(\omega))_+ + ( \mathbf{\hat{S}}_-^{-1}(\omega)  \mathbf{\hat{T}}(\omega))_-.
    \label{eq:Afac}
\end{equation}

Here, the multiplicative factorization of $ \mathbf{\hat{S}}(\omega)$ is solved numerically, based on the work done by \citet{daniele2007}. With these two factorizations mentioned above, the solution of Eq. (\ref{eq:WHf}) is given by equations (\ref{eq:WHs1}) and (\ref{eq:WHs2}):

\begin{equation}
   \mathbf{\hat{W}}_+(\omega) = \mathbf{\hat{S}}_+^{-1}(\omega)( \mathbf{\hat{S}}_-^{-1}(\omega)  \mathbf{\hat{T}}(\omega))_+, 
   \label{eq:WHs1}
\end{equation}

\begin{equation}
   \mathbf{\hat{W}}_-(\omega) = \mathbf{\hat{S}}_-(\omega)( \mathbf{\hat{S}}_-^{-1}(\omega)  \mathbf{\hat{T}}(\omega))_-,
    \label{eq:WHs2}
\end{equation}

\noindent $\mathbf{\hat{W}}_+$, which is regular in the upper-half complex plane, it is the term related to causal control.

\subsection{Control law design based on the Wiener-Hopf approach}

The Wiener-Hopf based control is considered for a linear dynamical system. An optimal control for such system can be obtained by means of a quadratic cost functional that expresses the control cost and the cost of state deviation from original condition, where this functional is minimized with respect to the control kernel. For simplicity, we first consider a case without feedback contamination i.e., the effect of $u$ on $y$ is neglected. Causality can be enforced by including Lagrange multipliers to the quadratic cost functional,

\begin{equation}
  J  = \int ^{\infty} _ {-\infty}\left(\langle{{{Q}{z^{*}}}{{z}}}
  + {Ru_+^{*}}{u_+} \rangle \right) dt + \int_{-\infty}^{\infty}[{{\Gamma_+}}(t){\Lambda_-}(t)]dt + \int_{-\infty}^{\infty}[{{\Gamma_+^*}}(t){\Lambda_-^*}(t)]dt,
  \label{eq:funcWH}
\end{equation}

\noindent where the $*$ superscript denotes the complex conjugate, ${z}$ is the target for the control problem, $Q$ is a weighting of the state deviation from original condition, $R$ is a control penalisation, ${\Lambda_-}$ is a Lagrange multiplier, $\Gamma$ is the control kernel, and ${u}$ is the actuation signal defined in frequency domain by eq. (\ref{eq:Kf_IFFC}). As $\Lambda$ is anticausal, the two last integrals in eq. (\ref{eq:funcWH}) force $\Gamma$ to be causal.

\begin{equation}
       \mathbf{\hat{u}}(\omega) = \mathbf{\hat{\Gamma}}(\omega)\mathbf{\hat{y}}(\omega) 
       \label{eq:Kf_IFFC}
    \end{equation}

\noindent In the time domain, the actuation is given by the following convolution,

\begin{equation}
       \mathbf{u}(t) = \int ^{\infty} _{-\infty} \mathbf{\Gamma}(\tau)\mathbf{y}(t-\tau)d\tau,
       \label{eq:kt_IFFC}
 \end{equation}

 \noindent which can lead to a non-causal kernel if the causality restriction is not imposed by the Lagrange multipliers. When Eq. (\ref{eq:funcWH}) is solved for the optimal solutions in the frequency domain, the following Wiener-Hopf equation is obtained:

\begin{equation}
  {\hat{H}_l}{\hat{\Gamma}}_+{\hat{S}_{yy}}  + {\hat{\Lambda}}_- = {\hat{H}_r}{\hat{S}_{yz}},
  \label{eq:WHeqctrl}
\end{equation}

\noindent where:

\begin{equation}
  {\hat{H}_l}(\omega)  = {\hat{G}_{uz}}^{*}(\omega) {Q}{\hat{G}_{uz}}(\omega)+{R},
  \end{equation}
  
\begin{equation}
    {\hat{H}_r}(\omega) = -{\hat{G}_{uz}}^{*}(\omega) {Q},
\end{equation}


\noindent ${\hat{S}_{yy}}(\omega)$ is the power spectral density (PSD) of the input signal ${y}$ and ${\hat{S}_{yz}}(\omega)$ is the cross-spectral density (CSD) between the observation signal ${y}$ and the target signal ${z}$ without actuation; ${G_{uz}}$ is the transfer function from the control signal ${u}$ to the output ${z}$. Applying the factorization presented earlier, the following expression is obtained for ${\hat{\Gamma}}_+$,

\begin{equation}
  {\hat{\Gamma}}_+ = {\hat{S}_{yy \hspace{0.075cm}+}^{-1}}({\hat{S}_{yy \hspace{0.075cm}-}^{-1}} {\hat{S}_{yz}} {\hat{H}_{r}} {\hat{H}_{l\hspace{0.075cm}-}^{-1}})_+ {\hat{H}_{l\hspace{0.075cm}+}^{-1}}.
  \label{eq:WHeqctrld}
\end{equation}

The expression above is thus used for obtaining the optimal causal control law under the Wiener-Hopf formalism \citep{Noble_1958, Martinelli2009, Martini2022}. 

In the present system there is a feedback contamination of the sensors, $y$, by the actuators, $u$, due to their proximity in the experimental setup. To account for this effects, and avoid appearance of a Larson effect \citep{Samfilippo2013}, the Wiener-Hopf kernel is modified as shown below  \citep{Martini2022}:   

\begin{equation}
  {\hat{\Gamma'}}_+ = ({{I}}+{\hat{\Gamma}}{\hat{G}_{uy}})^{-1},
\end{equation}

\noindent where ${\hat{G}_{uy}}$ is the transfer function between the actuation signal and the microphone readings.

\subsection{Control law design based on the inverse feed-forward control method}

In the wave-cancelling approach, here also referred to as inverse feed-forward control method, the strategy for control involves a direct superposition of the predicted uncontrolled disturbance at the objective position with the actuation. The IFFC control law can be obtained by minimizing a quadratic functional cost, similar to what is shown in Eq. (\ref{eq:funcWH}), but without the constrains imposed to obtain a causal kernel, i.e., without the Lagrange multiplier terms, which yields the following equation \citep{Sasaki2018, Brito2021}:

 \begin{equation}
  {\hat{\Gamma}} = \frac{({\hat{G}_{uz}})^{*}{Q}{\hat{G}_{yz}}}{({{\hat{G}_{uz}})^{*}}{Q}({\hat{G}_{uz}})+{R}},
  \label{eq:CtrlIFFCeqy_NoFB}
\end{equation}

\noindent where $\hat{G}_{yz}$ is the transfer function between the microphones $y$ and the control target $z$, which is simply ${\hat{S}_{yz}}/{\hat{S}_{yy}}$. Once the inverse Fourier transform of the IFFC kernel is taken, the non-causal part is neglected, i.e, we consider $\Gamma(\tau <0) = 0$ in order to be able to apply the IFFC kernel in experiments.

In order to account for the contamination of the sensor readings caused by the actuators, it is also necessary to take into consideration the transfer function between the actuators and the microphones ($G_{uy}$), in this case the IFFC kernel can be obtained as

\begin{equation}
  {\hat{\Gamma}} = \frac{({\hat{G}_{uz}} -{\hat{G}_{uy}}{\hat{G}_{yz}})^{*}{Q}{\hat{G}_{yz}}}{({{\hat{G}_{uz}} -{\hat{G}_{uy}}{\hat{G}_{yz}})^{*}}{Q}({\hat{G}_{uz}} -{\hat{G}_{uy}}{\hat{G}_{yz}})+{R}}.
  \label{eq:CtrlIFFCeq}
\end{equation}

\subsection{Further considerations about the control laws}

For the present work the CSDs and PSDs were obtained using the Welch method, where the Hanning window has been considered with 218 blocks and 50\% overlap. Since we have considered a sampling rate of $10$ kHz ($St = 29.9$)and measurements of $30$ s to identify the transfer functions, we obtained a frequency resolution of $11$ Hz (St = 0.03), with a total of $2730$ data points per block. 

To remove some of the undesired noises from the kernel, a zero-phase band-pass filter was applied to the raw data before the calculation of the transfer functions. The application of a filter was important because even very small amplitude noise in the kernel could result in a rapidly increase of the actuation amplitudes due to the acoustic feedback contamination. A frequency-based control penalisation was used for the control kernels, wherein large penalisations were applied outside the desired frequency-band, defined previously. This helps avoiding undesired frequencies that can lead to feedback contamination and it also regularizes the problem, avoiding divisions by zero (or by values close to zero) in eqs. (\ref{eq:WHeqctrld}) and (\ref{eq:CtrlIFFCeqy_NoFB}). To further regularize the problem, a noise term of constant value $(1\cdot10^{-4})$ was added to the PSDs of $u$ and $y$. For the case where a forcing is applied up to Strouhal number of $0.85$, better results where obtained using a constant value of $(1\cdot10^{-3})$.

\begin{figure}
\includegraphics[width=1\textwidth]{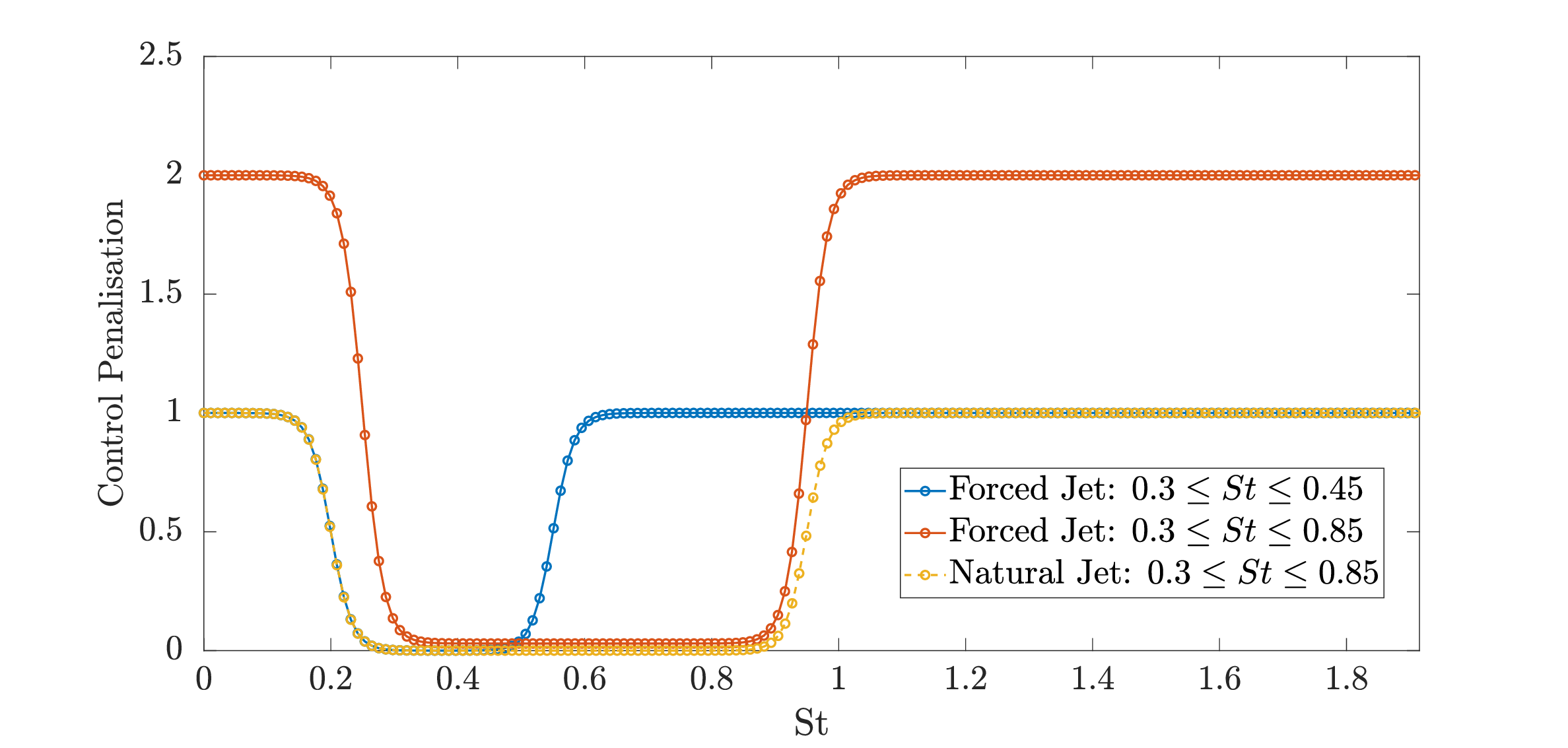}
\caption{Control penalisation as function of the frequency.}
\label{fig:Cont_penal}
\end{figure}

 The expressions used to obtain the control penalisation are shown in Eqs. (\ref{eq:Cont_penalIg1}-\ref{eq:Cont_penalIg}),

\begin{equation}
R_1 = R_{ctrl} + a(R_{noise}-R_{ctrl})(\tanh(f + f_{c1})-\tanh(f - f_{c1})),
\label{eq:Cont_penalIg1}
\end{equation}

\begin{equation}
R_2 = R_{ctrl} + a(R_{ctrl} - R_{noise})(\tanh(f + f_{c2})-\tanh(f - f_{c2})),
\label{eq:Cont_penalIg2}
\end{equation}

\begin{equation}
R= 
\begin{cases}
     R_1,& \text{if } f_c \leq (f_{c1} + f_{c2})/2 \\
    R_2,          & {\mathrm{otherwise}}
\end{cases}
\label{eq:Cont_penalIg}
\end{equation}

\noindent and are illustrated in fig. \ref{fig:Cont_penal}. Where $f$ is a frequency vector; $f_{c1}$ and $f_{c2}$ are cut-off frequencies; $R_{ctrl}$ defines the penalisation within the region of interest; $R_{noise}$ is the penalisation out of the region of interest; We have used $a = 30$, which is a parameter that defines how fast and smooth is the transition from $R_{ctrl}$ to $R_{noise}$. The parameters used in each case are summarized in table \ref{tab:parameters}.

\begin{table}
  \begin{center}
\def~{\hphantom{0}}
  \begin{tabular}{lccccc}
      & Frequency range & $f_{c1}$   & $f_{c2}$ & $R_{ctrl}$  & $R_{noise}$ \\[3pt]
   Forced jet & ($0.3 \leq St \leq 0.45$) & 0.20 & 0.55 & $1\cdot10^{-5}$ & 1\\
   Forced jet & ($0.3 \leq St \leq 0.85$) & 0.25 & 0.95 & $3\cdot10^{-2}$ & 2\\
   Natural jet & ($0.3 \leq St \leq 0.85$)& 0.20 & 0.95 & $1\cdot10^{-5}$ & 1\\
  \end{tabular}
  \caption{Parameters considered for the control penalisation.}
  \label{tab:parameters}
  \end{center}
\end{table}


\begin{figure}
\includegraphics[width=1\textwidth]{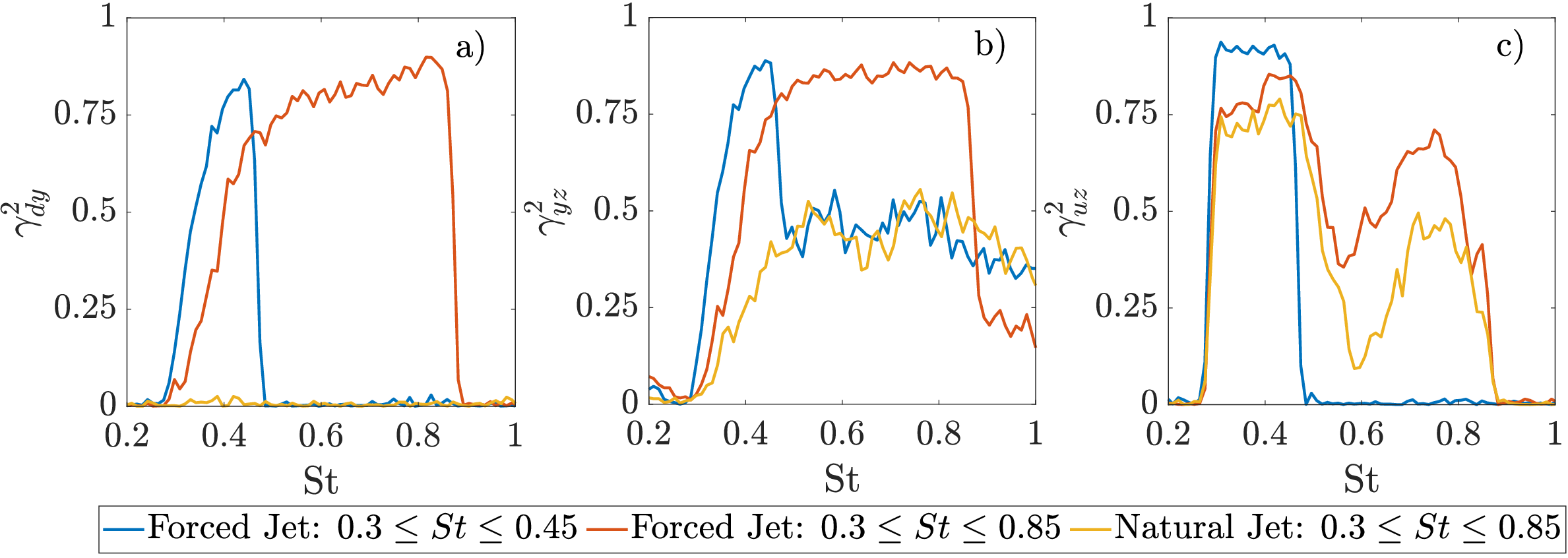}
\caption{Coherence between: (a) forcing, $d$, and reference signal, $y$; (b) reference signal, $y$, and control target, $z$, and; (c) actuation signal, $u$, and control target, $z$.}
\label{fig:Cohe_3sigs_03_045}
\end{figure}

The functions related to the actuation (${\hat{G}_{uy}}$ and ${\hat{G}_{uz}}$) were obtained with the forcing off, while $S_yy$ and $S_yz$ were obtained with the forcing (for the forced jet cases) on and the actuation off. The coherence obtained between the sensors and actuators are shown in fig. \ref{fig:Cohe_3sigs_03_045}. High coherence levels are required for the linear control approaches employed used here. 

As can be observed in the forced jet case, with the increase of the bandwidth significant lower coherence levels, between the reference sensors and the target, were observed for Strouhal numbers below $0.45$. For $St = 0.37$, for example, $\gamma_{yz}$ dropped by about $50\%$ with the larger forcing bandwidth. For the forced jet with $0.3 \leq St \leq 0.45$, $\gamma_{yz}$ can also be considered relatively low for $St = 0.34$ and lower, where $\gamma_{yz}$ was below $0.6$. This lower coherences can be explained either by the microphones not being able to properly identify the $m = 0$ structures, e.g., due to azimuthal aliasing, or due to decoherence of the wavepacket along the jet. Unlike transitional flows, characterised by small disturbances, we deal here with a turbulent jet, with coherent structures that nonetheless display an intrinsic coherence decay, imposing additional challenges for flow control. Nevertheless, for the forced jet cases, the coherence between sensors were mostly above $0.8$ for the frequencies of interest. Regarding the natural jet case, low coherence levels were obtained for the entire frequency range considered for control, around $0.5$, which makes the control of natural turbulent jets a much more challenging task. The coherence between the actuation signal and the hot wire was higher than $0.85$ for the region of interest, mostly above $0.9$, when considering the forced jet case with $0.3 \leq St \leq 0.45$. However, as the actuation bandwidth increases, lower coherence levels are obtained, which was found to be a limitation of the actuation system. Differences in coherence levels observed between the forced and unforced cases ($0.3 \leq St \leq 0.85$) can be partially explained by differences in the amplitudes of the actuation signals when identifying the transfer functions, but it was observed that this does not have a significant impact in the obtained kernel.

%% file: sec_Results.tex
 \section{Results}\label{PrelResults}

\subsection{Control laws}

\begin{figure}
\includegraphics[width=1\textwidth]{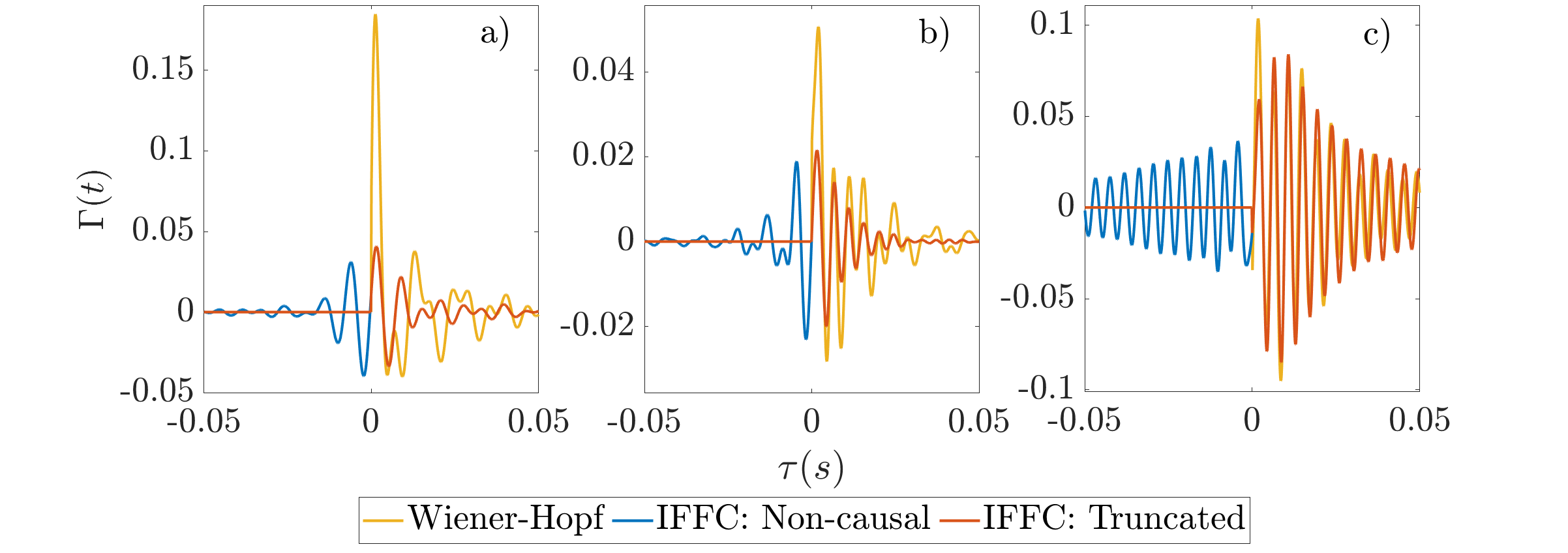}
\caption{Control kernels in time domain for the forced jets (a) $0.3 \leq St \leq 0.45$ (b) $0.3 \leq St \leq 0.85$, and (c) for the natural jet (actuation bandwidth = $0.3 \leq St \leq 0.85$).}
\label{fig:kernelsy_03_045}
\end{figure}

\begin{figure}
\includegraphics[width=1\textwidth]{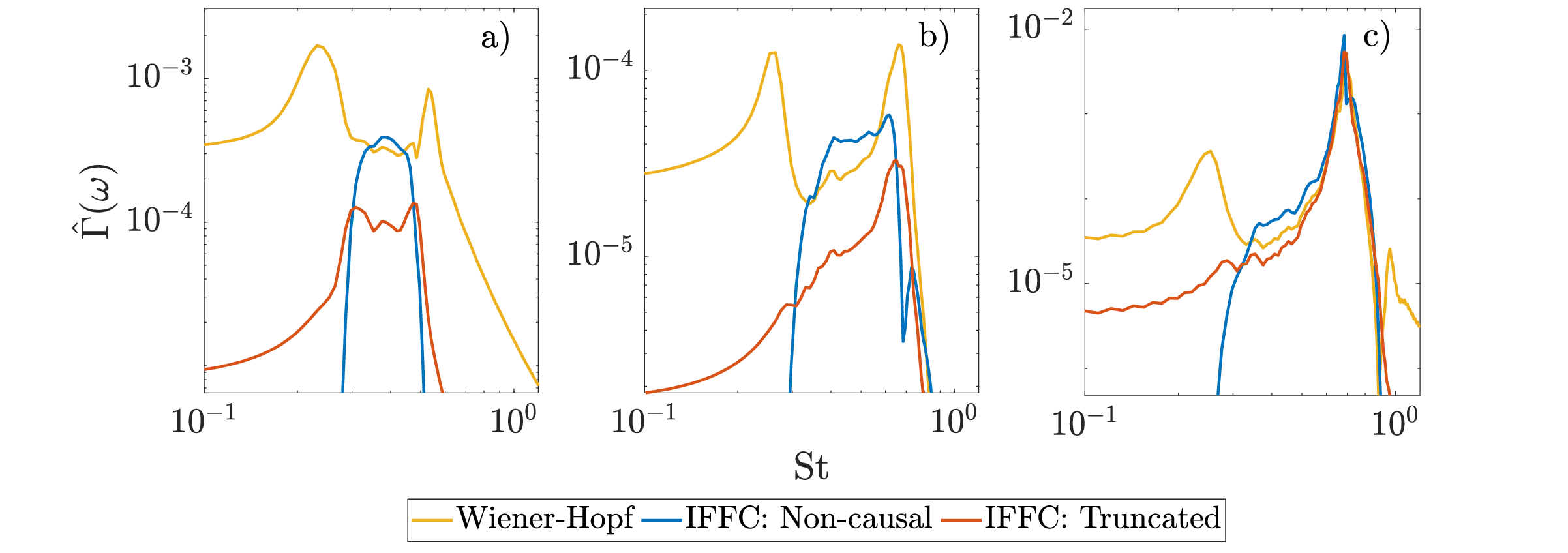}
\caption{Control kernels in frequency domain for the forced jets (a) $0.3 \leq St \leq 0.45$, (b) $0.3 \leq St \leq 0.85$, and (c) for the natural jet (actuation bandwidth = $0.3 \leq St \leq 0.85$).}
\label{fig:kernels_freq}
\end{figure}

The kernels obtained with the inverse feed-forward control (IFFC) method and the Wiener-Hopf technique, which were discussed in the previous section, are compared in fig. \ref{fig:kernelsy_03_045} and \ref{fig:kernels_freq}.

In practical applications, it is unfeasible to apply the non-causal part of the IFFC kernel ($\tau < 0$) shown in figure \ref{fig:kernelsy_03_045}. For that reason, the non-causal kernel needs to be truncated to its causal part, i.e., we set $\Gamma(\tau < 0) = 0$, which reduces the performance of the controller. The Wiener-Hopf approach provides an optimal truncation strategy, minimizing the performance loss imposed by the causality constrain. The kernels obtained for the natural jet contain oscillations that do not decay to zero, as would be expected \citep{herve2012, fabbiane2015, Borggaard2016, Karban2023}, in this case they display a harmonic-like signature, due to the peak in $\Gamma({\omega})$ around St=0.65. This is a way for the controller to compensate the low actuator-objective coherence (fig. \ref{fig:Cohe_3sigs_03_045}c). This tonal behaviour results in the slow decay $\Gamma(t)$. For the forced jet cases, as the flow is driven by a stochastic signal, the coherent time-lengths are dominated by the forcing signal, not by the intrinsic flow dynamics. 

Wiener-Hopf kernels present higher amplitudes at $\tau \approx 0$, which is seen as a compensation to the non-causality observed in the IFFC kernels, as observed in \citet{Martini2022} and \citet{Audiffred2023}. When observing the spectra of the kernels, figure \ref{fig:kernels_freq}, one may notice that the energy content of the Wiener-Hopf kernels, for the forced cases, is similar to the non-causal ones in the region we are aiming the control, while the truncated non-causal solution has considerably lower energy content in this same region, about $70\%$ lower for both forced jet conditions. On the other hand, some undesired peaks out of the region of interest appeared for the Wiener-Hopf kernels. That likely occurs because we are exciting the jet flow with a band-limited stochastic forcing instead of a white noise. Additionally, the raw data is filtered in order to remove some of the noise present in the region we are not aiming the control, as mentioned earlier. The higher control penalisation for this same region prevents that those peaks extend for a larger bandwidth. Those peaks could be attenuated by modifying the cut-off frequencies in equations \ref{eq:Cont_penalIg1} and \ref{eq:Cont_penalIg2}, but that would negatively affect the performance of the kernel.

\subsection{Control results}

\begin{figure}
\includegraphics[width=1\textwidth]{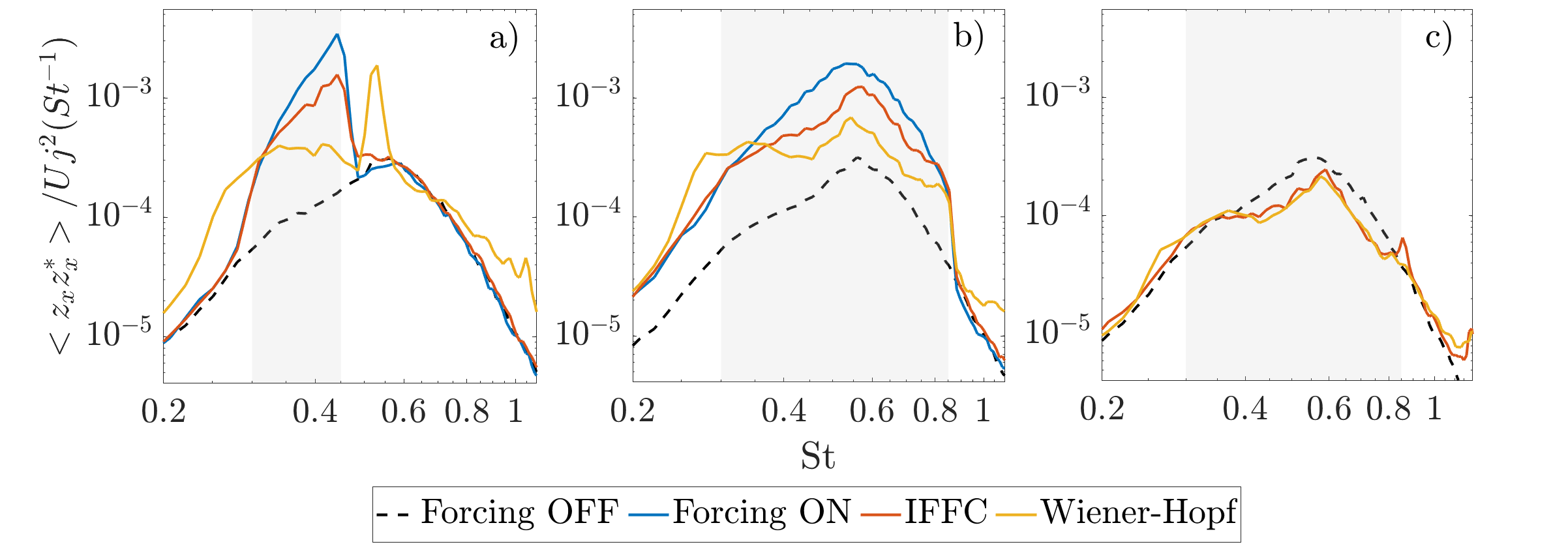}
\caption{Spectra of the uncontrolled and controlled turbulent jet flow comparing the Wiener-Hopf and IFFC at the control target location, for the kernels obtained for the forced jets (a) $0.3 \leq St \leq 0.45$, (b) $0.3 \leq St \leq 0.85$ and (c) unforced jet, with the control aimed for $0.3 \leq St \leq 0.85$.}
\label{fig:specCtrl_03_045}
\end{figure}

The control results obtained with the kernels presented above are shown in fig. \ref{fig:specCtrl_03_045}, which displays the average PSDs of the objective, obtained from $3$ measurements of $15$ s each, with a sampling frequency of $10$ kHz. In \citet{AudiffredAIAA} we demonstrated, using this same data set, the control authority, i.e., that the reductions obtained are due to reactive control, not due to changes in the flow dynamics. The same has been already demonstrated by \citet{Maia2021}. 

For the forced jet cases, a significantly better performance is observed with the Wiener-Hopf kernel in comparison with the IFFC law. Here, the IFFC kernel had a worse performance than reported by \citet{Maia2021}. In the present study we noticed that the actuator-objective transfer function, $G_{uz}$, was significantly different from that of \citet{Maia2021}. This suggests a difference in actuator placement, or in the intrinsic mechanical characteristics of the actuation system. In any case, the Wiener-Hopf kernel still performed better than what is reported by \citet{Maia2021}. For the case where a forcing is applied for  $0.3 \leq St \leq 0.45$, a large peak is observed for the Wiener-Hopf results at Strouhal number around $0.52$. Although this is undesired, the peak appears outside the frequency range we are aiming to control. This undesired peak resulted from one of the peaks observed in the Wiener-Hopf control kernel. For the case with larger forcing bandwidth, we have observed a considerable increase in the broadband noise outside the frequency range of the excitation signal, which has also been observed in earlier studies \citep{moore1977, Maia2021}. For the natural jet, similar results were obtained with both control methods considered here, with an attenuation of power spectral density of about $40\%$, comparing the uncontrolled and controlled spectra, for the St range of higher amplification of the axisymmetric pressure mode. 

\subsection{Downstream persistence of control effects}
 
\begin{figure}
\includegraphics[width=0.95\textwidth]{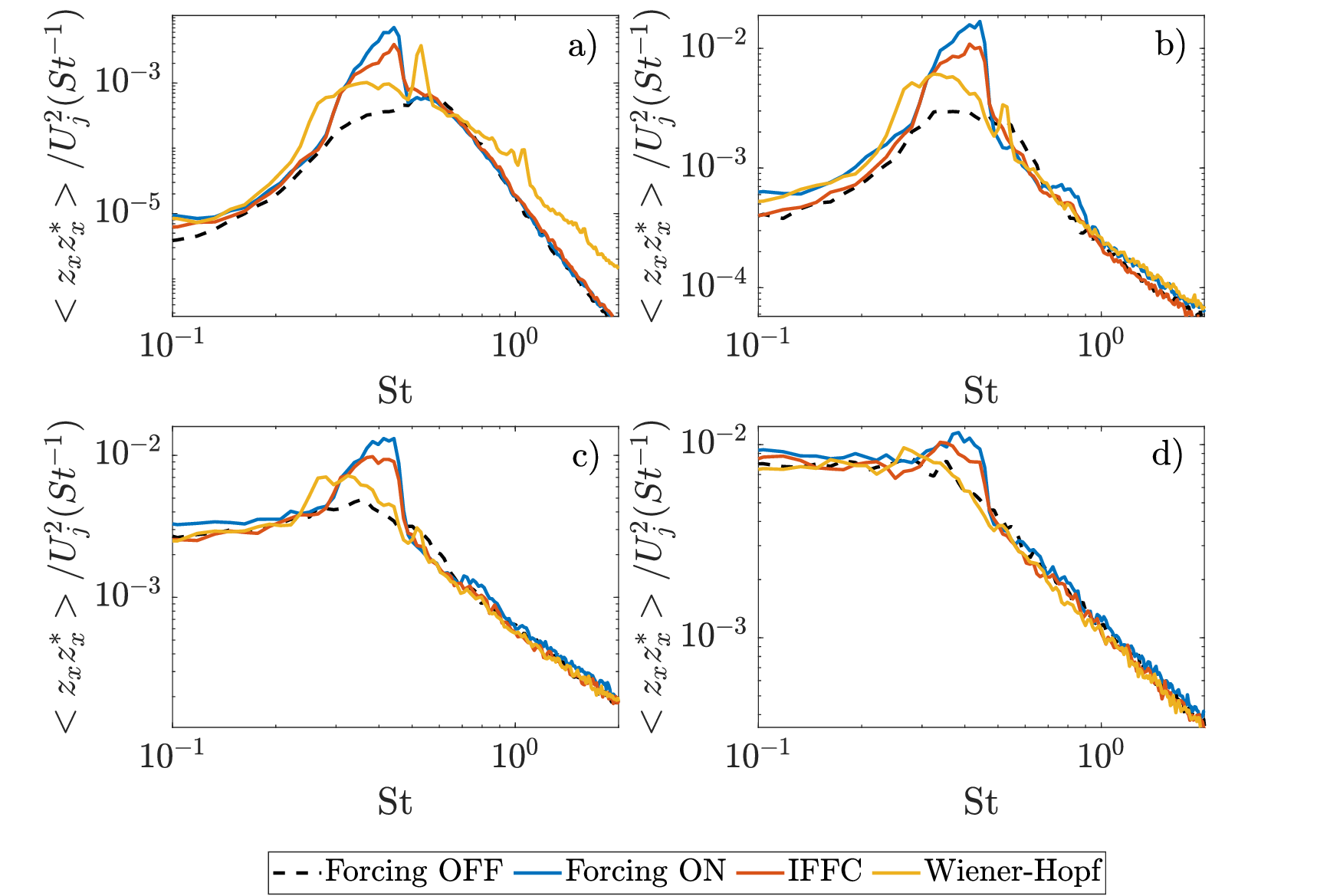}
\caption{Spectra of the uncontrolled and controlled forced jet flow ($0.3\leq St\leq0.45$) at different positions in the streamwise direction: (a) $x/D = 2.5$, (b) $x/D = 5$, (c) $x/D = 6$, (d) $x/D = 7$.}
\label{fig:specCtrl_03_045VarPos}
\end{figure}

\begin{figure}
\includegraphics[width=0.95\textwidth]{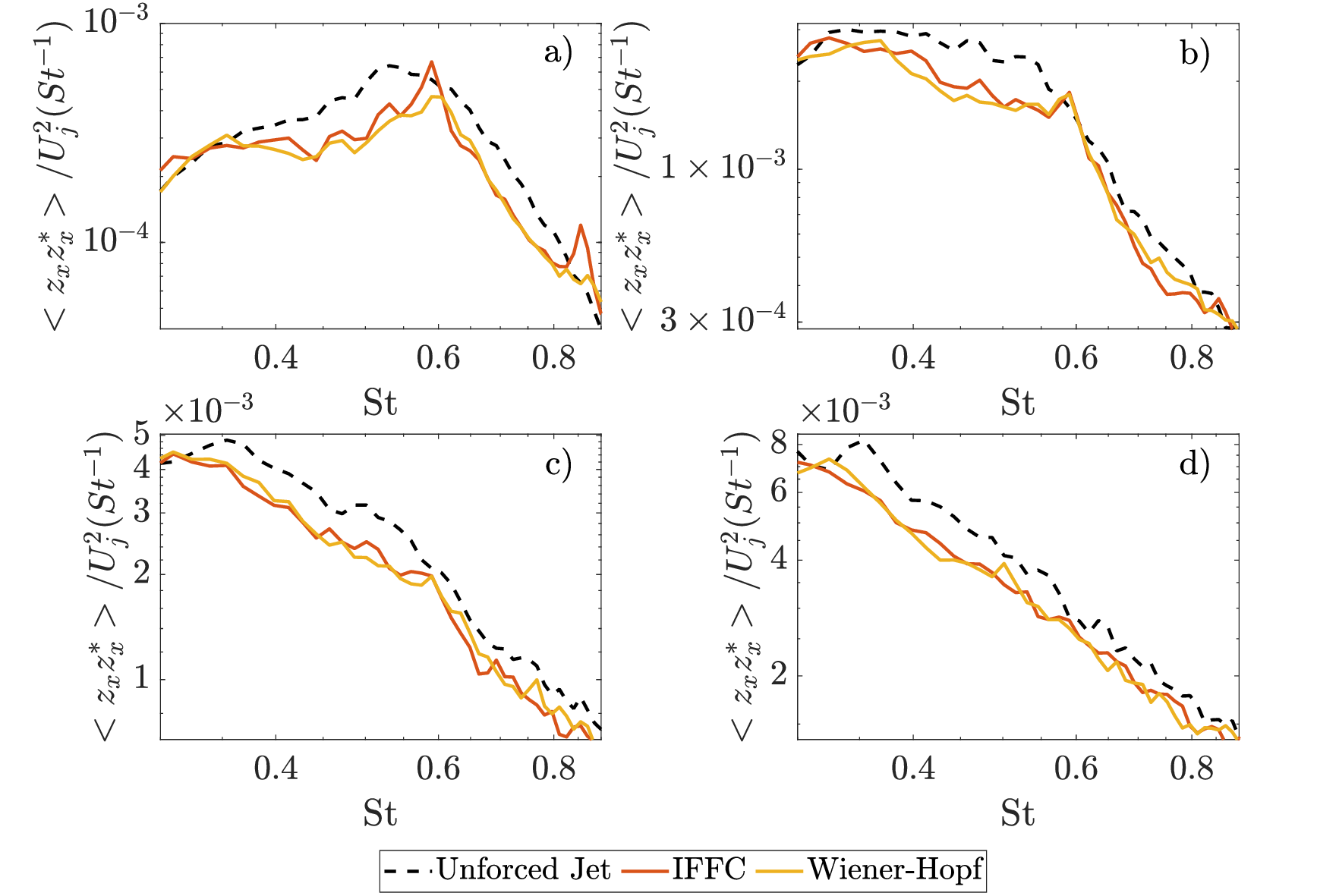}
\caption{Spectra of the uncontrolled and controlled natural jet flow at different positions in the streamwise direction: (a) $x/D = 2.5$, (b) $x/D = 5$, (c) $x/D = 6$, (d) $x/D = 7$.}
\label{fig:specCtrl_03_085NatVarPos}
\end{figure}

We also verified whether the control effect persisted downstream of the target position for the forced jet case with $0.3 \leq St \leq 0.45$ and for the natural jet. A significant attenuation of the signals could still be observed until the streamwise position $x/D = 7$, as shown in figs. \ref{fig:specCtrl_03_045VarPos} and \ref{fig:specCtrl_03_085NatVarPos}. For the forced case, we can observe that the Wiener-Hopf approach had a more positive impact for frequencies closer to the upper limit of the targeted frequency range compared to the IFFC approach, such that no forcing effect is visible for the Wiener-Hopf results at $x/D = 7$. For the natural jet, reductions of about $25 \%$ can be observed in the velocity spectra at $x/D = 7$, showing that control benefits are not restricted to the target location

\subsection{Improvement on sensor placement}

\begin{figure}
\includegraphics[width=0.95\textwidth]{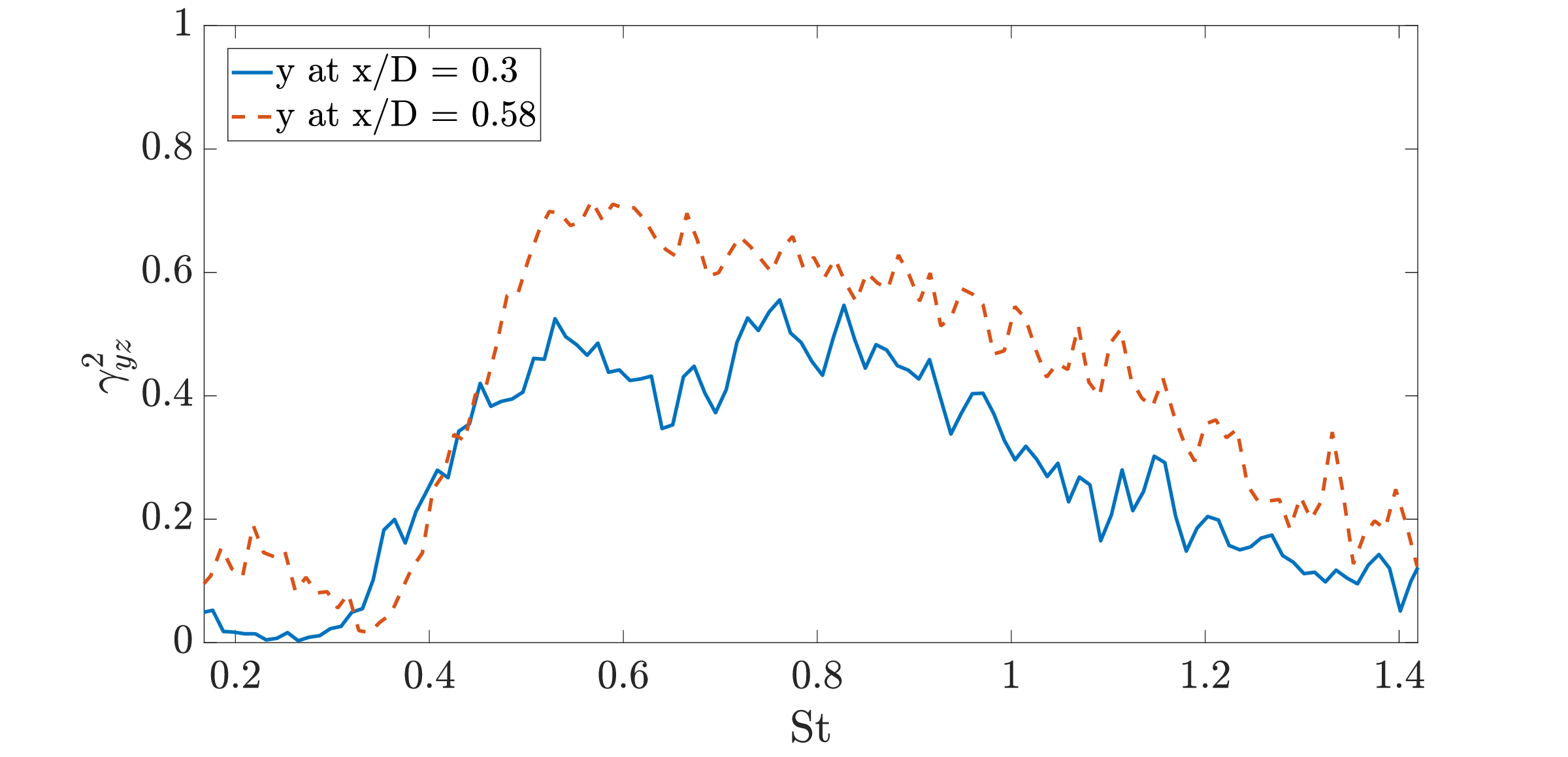}
\caption{Coherence between the microphones and the hot wire, comparing the case with the microphones at $x/D = 0.3$ and at $x/D = 0.58$.}
\label{fig:coherence_Y14mm}
\end{figure}

As observed in fig. \ref{fig:Cohe_3sigs_03_045}, the coherence between sensors drops significantly in the absence of a forcing applied to the jet. Since the amplitude of wavepackets increases as they are convected by the jet, it becomes easier to detect them. Thus, we considered positioning the array of microphones at a slightly  downstream position, at $x/D = 0.58$, remaining nonetheless close to the nozzle. With this, we also changed the radial position of the sensors, moving them to $r/D =0.6$, due to the thicker shear layer at that position. A comparison between the coherence between the target sensor $z$ and the two positions of $y$ sensors is shown in fig. \ref{fig:coherence_Y14mm}, confirming that the downstream $y$-sensor placement leads to higher coherence and thus better estimation capabilities.   

\begin{figure}
\includegraphics[width=0.95\textwidth]{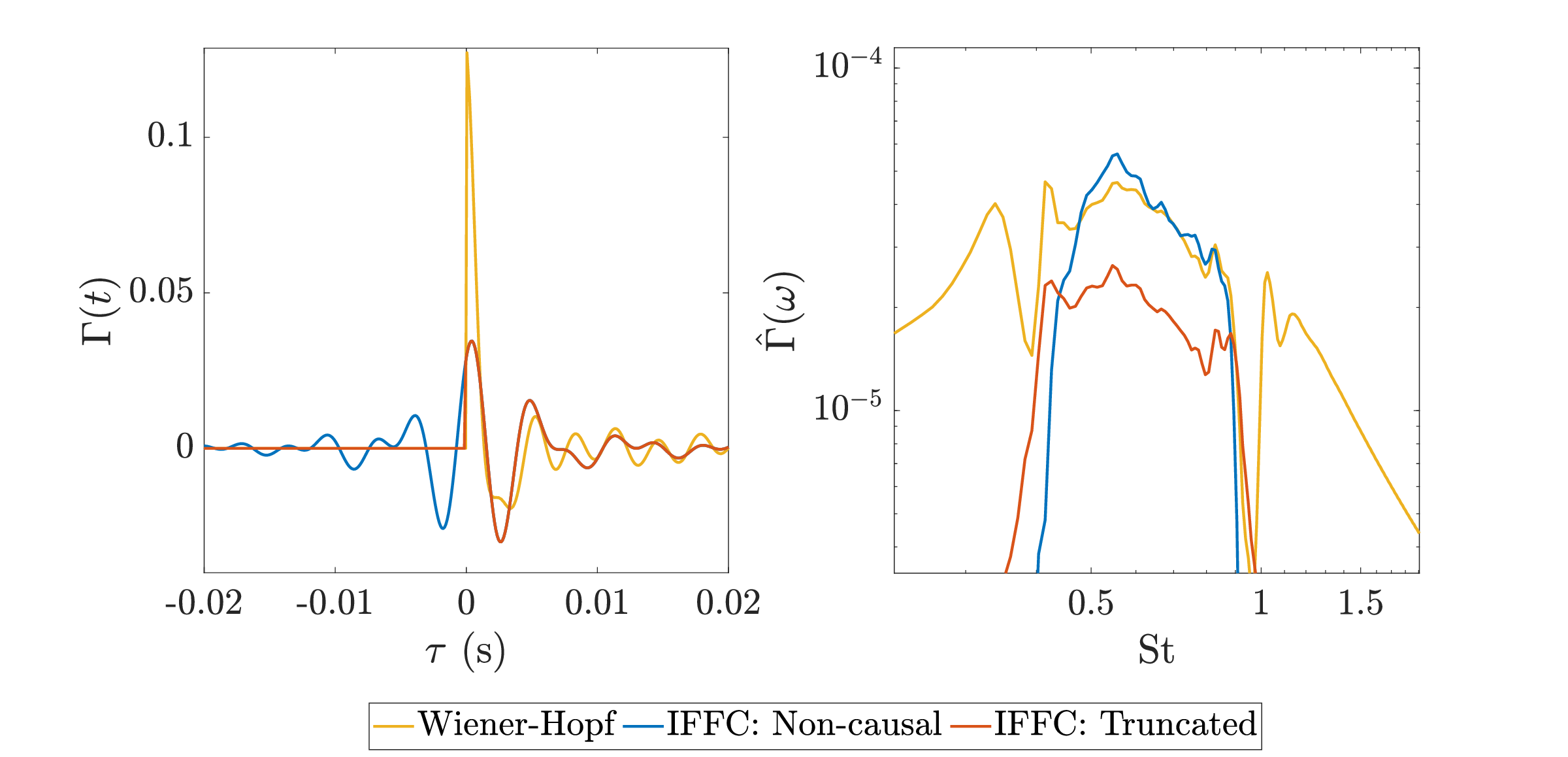}
\caption{Control kernels of the unforced jet with the sensors farther downstream, in (a) time domain and (b) frequency domain.}
\label{fig:kernels_Y14mm}
\end{figure}

Since the coherence is still low for $St<0.45$, we decided to restrict the actuation signal to the range $0.45 \leq St \leq 0.9$. The kernels for these new conditions are shown in fig.\ref{fig:kernels_Y14mm}. With this higher coherence, the obtained kernels became less oscillatory than the ones presented in fig. \ref{fig:kernelsy_03_045}, and the kernel signal has a stronger decay over time, as usual in flow control applications \citep{herve2012, fabbiane2015, Borggaard2016, Karban2023}.

\begin{figure}
\includegraphics[width=0.95\textwidth]{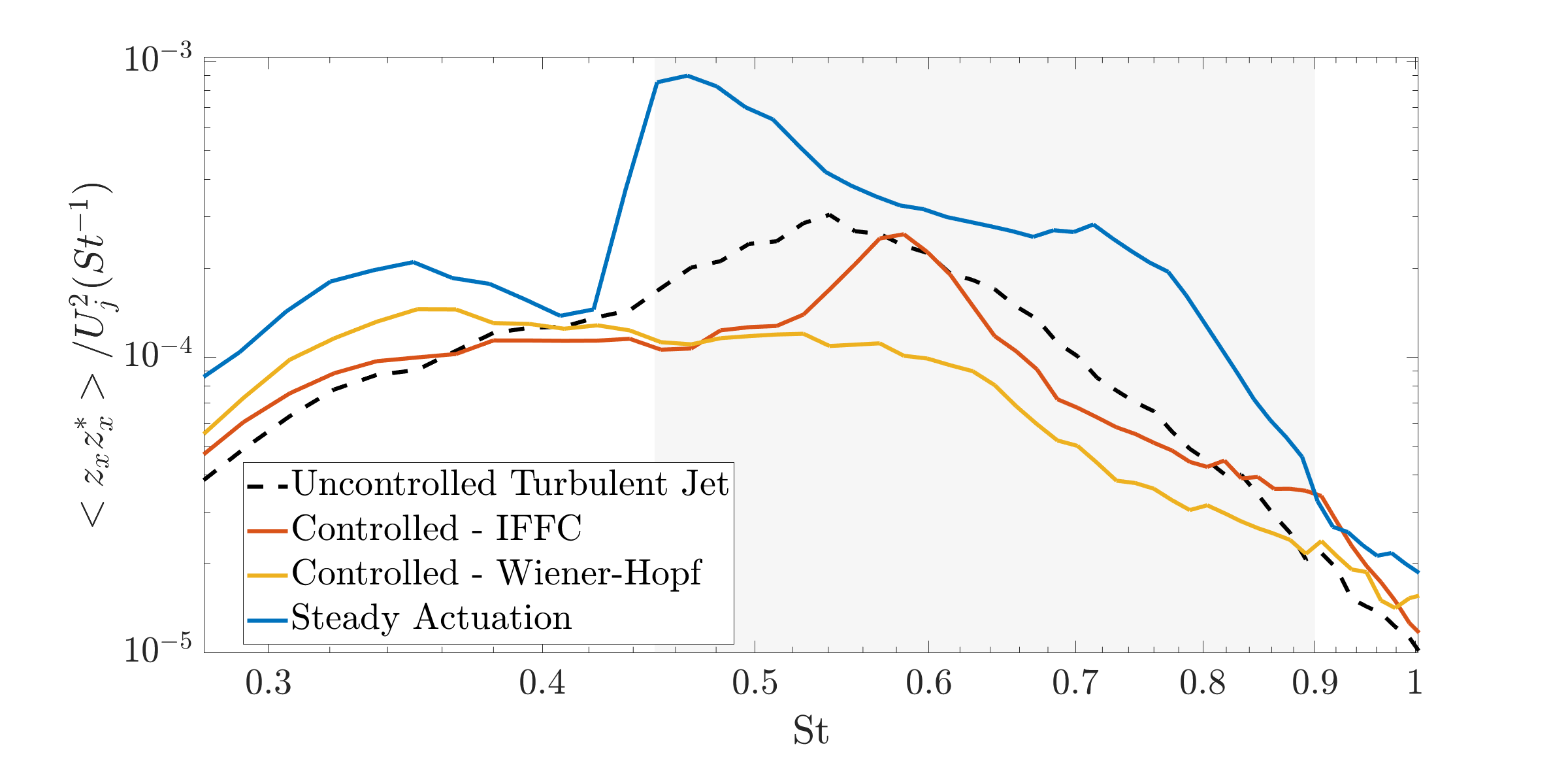}
\caption{Spectra of the uncontrolled and controlled turbulent jet flow comparing the
Wiener-Hopf and IFFC at the control target location.}
\label{fig:Szz_Y14mm_pos2}
\end{figure}

\begin{figure}
\includegraphics[width=0.95\textwidth]{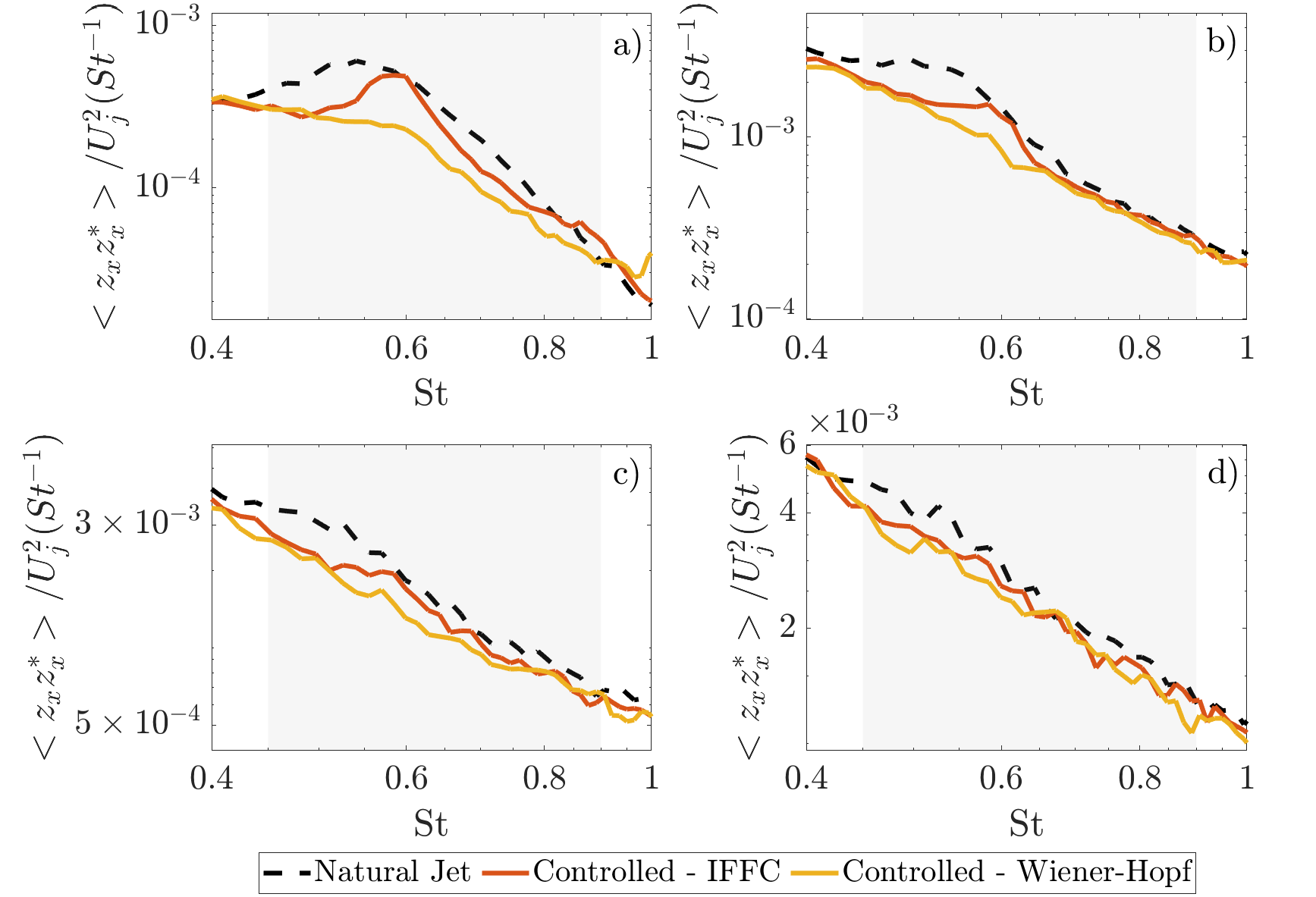}
\caption{Spectra of the uncontrolled and controlled turbulent jet flow (for $y$ at $x/D$ = 0.58), at different positions in the streamwise direction: (a) $x/D = 2.5$, (b) $x/D = 5$, (c) $x/D = 6$, (d) $x/D = 7$.}
\label{fig:Szz_Y14mm_posVar}
\end{figure}

The spectra of the uncontrolled and controlled signals obtained with these new kernels are presented in fig. \ref{fig:Szz_Y14mm_pos2} for the objective position and in fig. \ref{fig:Szz_Y14mm_posVar} for measurements taken farther downstream. With the Wiener-Hopf kernel, at the objective position, reductions of more than $60\%$ of power spectral density could be obtained for the most amplified frequencies. Attenuation close to that could also be obtained with the IFFC kernel, however, a much worse performance is observed at Strouhal number around $0.58$. We also considered here a steady actuation case, where the actuators provide a band-limited stochastic signal with the same frequency range target by the reactive control, from where can be noticed that a simply steady actuation cannot guarantee the attenuation of the wavepackets, on the contrary, we had a condition where it amplifies the coherent structures.

When the reference sensors get closer to the actuators, the non-causal part of IFFC kernels increases, which deteriorates the performance of the truncated kernel, as reported in \citet{Martini2022} and by \citet{Audiffred2023}. The worse performance of the IFFC kernel at Strouhal number around $St = 0.58$ seems also to be related to the low coherence presented by the actuator for this same region (fig. \ref{fig:Cohe_3sigs_03_045}), where the Wiener-Hopf kernel was able to compensated that, most likely due to an overall stronger effort of the actuation.

\begin{figure}
\includegraphics[width=0.95\textwidth]{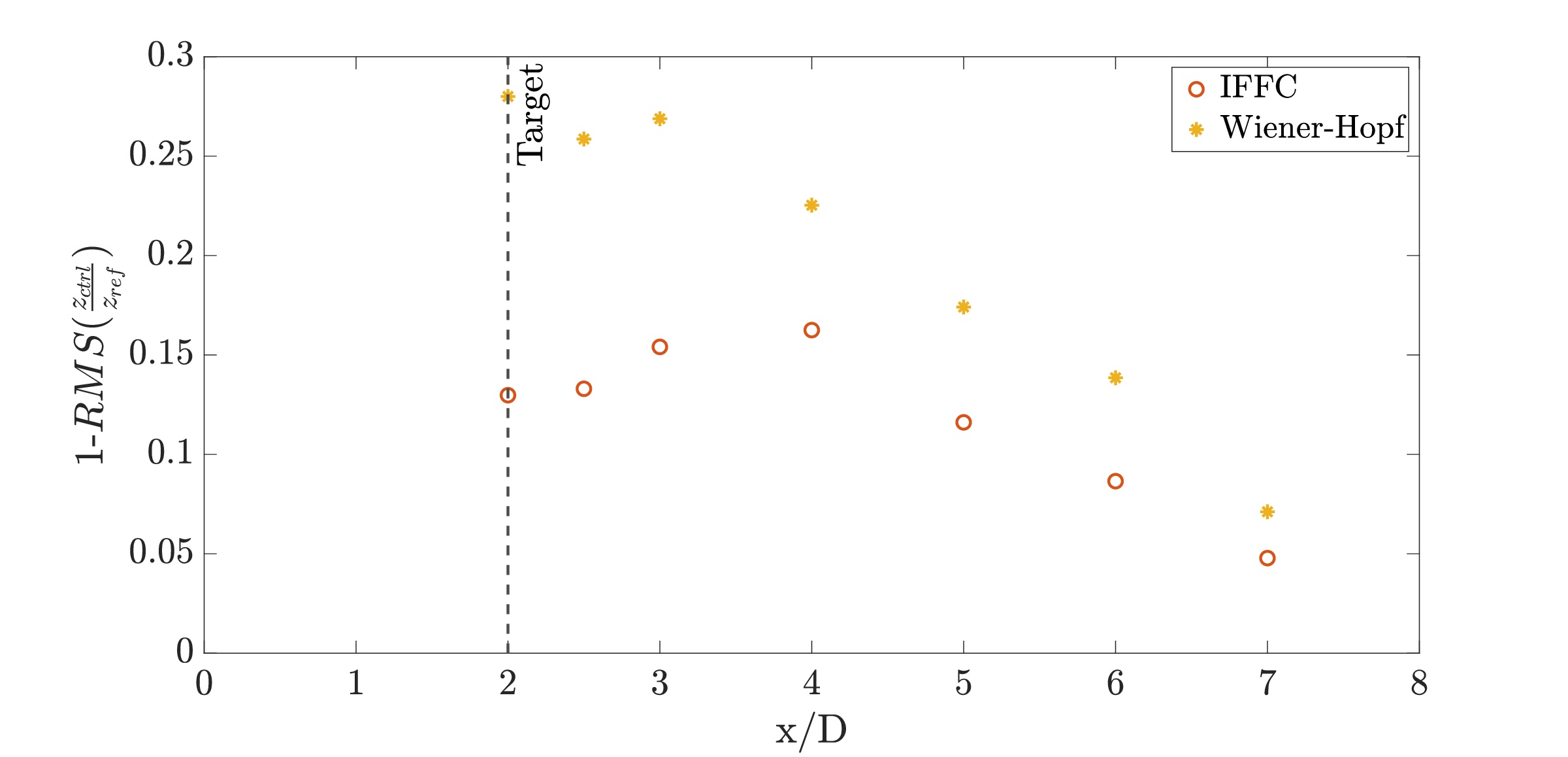}
\caption{Relative performance of the control kernels, in terms of the $RMS$ of velocity fluctuations of the controlled signals at different positions.}
\label{fig:Perf_rmsY14mm_posVar}
\end{figure}

We also evaluated the overall performance of the kernels, at different streamwise positions, in terms of the root mean square (RMS) of velocity fluctuations of the controlled and uncontrolled signals. The result of this, for the unforced case with the sensors farther downstream, is shown in fig. \ref{fig:Perf_rmsY14mm_posVar}. The Wiener-Hopf kernel could provide a reduction of $28\%$ of the RMS of the velocity fluctuations at the target location, which represents an overall reduction of $50\%$ if we consider the integral of the PSDs over the target frequency range, instead of the RMS signal. On the other hand, the IFFC approach provided a RMS reduction of $12\%$. The performance of the IFFC kernel reaches a peak at $x/D = 4$, with a RMS reduction of $16\%$. Farther downstream, it is observed a linear decay of the control effect.  

%% file: sec_Conclusion.tex
\section{Conclusion}\label{conclusion}

This work presented an experimental campaign on reactive control of turbulent jets, where a Wiener-Hopf approach was used to obtain the control kernels, while the inverse feed-forward control method was used as comparison. The control law aims at the attenuation of axisymmetric disturbances associated with Kelvin-Helmholtz wavepackets, known to be related to the peak subsonic jet noise \citep{PeterTim}. The Wiener-Hopf-based control improves on previous results reported by \citet{Maia2021}, for turbulent jets forced stochastically. Furthermore, we extend the control experiments to natural unforced jets; this is the first time, to the best of our knowledge, that reactive control is performed on natural turbulent jets, i.e. turbulent jets without any artificial forcing. Attenuation of about $60 \%$ in fluctuation energy was obtained in the unforced jet case. It was also observed that the reduction of the wavepackets persisted further downstream of the control target location, for both the forced and unforced cases.

This study was motivated by positive results obtained from other previous works, such as the use of the Wiener-Hopf technique for the application in flow control, both numerically \citep{Martini2022} and experimentally \citep{Audiffred2023}, and the experimental control of forced jets using the IFFC method \citep{Maia2021}. Although the power spectra and phase differences of forced wavepackets are similar to what is found in unforced turbulent jets, the control of a natural jet becomes more challenging due to lower coherence levels, and they are a better representation of real applications. Furthermore, despite difficulties with feedback contamination of sensor readings and poor actuator performance at large bandwidths, control was found to be successful in generating substantial energy attenuation.

The placement of sensors had an important influence in the control of the natural jet. Once sensors were moved from $x/D = 0.3$ to $x/D = 0.58$, better coherence levels with the downstream target sensor could be obtained, and with this, the far-field noise attenuation improved accordingly. Within this context, the Wiener-Hopf technique played a major role in order to deal with the non-causality observed in the IFFC kernels, leading to optimal control laws that were able to further reduce velocity fluctuations. Where we were able to reach a reduction of $28 \%$ of the velocity fluctuations in terms of RMS values, and a target frequency range-integrated energy reduction of $50\%$.

Although this control experiment was performed for a low Mach number, the control of a natural turbulent jet presented here is an important step towards the control of turbulent jets under real flight conditions. Furthermore, the same sound radiation mechanisms are observed for higher Reynolds and subsonic Mach numbers, with sound radiation for low polar angles dominated by the axissymmetric mode \citep{cavalieri2012}. Therefore, the control strategy adopted here, with an improvement of the experimental setup, could be extended to higher Mach numbers with further technological applications.